\documentclass[sigconf]{acmart}
\AtBeginDocument{%
  }

\copyrightyear{2023}
\acmYear{2023}
\setcopyright{acmlicensed}\acmConference[SIGIR '23]{Proceedings of the 46th International ACM SIGIR Conference on Research and Development in Information Retrieval}{July 23--27, 2023}{Taipei, Taiwan}
\acmBooktitle{Proceedings of the 46th International ACM SIGIR Conference on Research and Development in Information Retrieval (SIGIR '23), July 23--27, 2023, Taipei, Taiwan}
\acmPrice{15.00}
\acmDOI{10.1145/3539618.3591950}
\acmISBN{978-1-4503-9408-6/23/07}

\usepackage{multirow}
\begin{document}

\title{Attention-guided Multi-step Fusion: A Hierarchical Fusion Network for Multimodal Recommendation}


\author{Yan Zhou}
\affiliation{%
 \institution{Xidian University}
 \city{Xi’an}
 \state{Shaanxi}
 \country{China}}
\authornotemark[2]
\email{zhouyan123y@stu.xidian.edu.cn}

\author{Jie Guo}
\affiliation{%
 \institution{Xidian University}
 \city{Xi’an}
 \state{Shaanxi}
 \country{China}}
 \authornote{Corresponding Author.}
 \authornote{State Key Laboratory of Integrated Services Networks.}
\email{jguo@xidian.edu.cn}

\author{Hao Sun}
\affiliation{%
 \institution{Xidian University}
 \city{Xi’an}
 \state{Shaanxi}
 \country{China}}
\authornotemark[2]
\email{21011210029@stu.xidian.edu.cn}

\author{Bin	Song}
\affiliation{%
 \institution{Xidian University}
 \city{Xi’an}
 \state{Shaanxi}
 \country{China}}
\authornotemark[2]
\email{bsong@mail.xidian.edu.cn}

 \author{Fei Richard	Yu}
\affiliation{%
 \institution{Shenzhen University}
 \city{Shenzhen}
 \state{Guangdong}
 \country{China}}
 \authornote{Guangdong Laboratory of Artificial Intelligence and Digital Economy (SZ).}
\email{yufei@szu.edu.cn}

\renewcommand{\shortauthors}{Zhou et al.}

\begin{abstract}
The main idea of multimodal recommendation is the rational utilization of the item's multimodal information to improve the recommendation performance. Previous works directly integrate item multimodal features with item ID embeddings, ignoring the inherent semantic relations contained in the multimodal features. In this paper, we propose a novel and effective aTtention-guided Multi-step FUsion Network for multimodal recommendation, named TMFUN.
Specifically, our model first constructs modality feature graph and item feature graph to model the latent item-item semantic structures. Then, we use the attention module to identify inherent connections between user-item interaction data and multimodal data, evaluate the impact of multimodal data on different interactions, and achieve early-step fusion of item features. Furthermore, our model optimizes item representation through the attention-guided multi-step fusion strategy and contrastive learning to improve recommendation performance. The extensive experiments on three real-world datasets show that our model has superior performance compared to the state-of-the-art models.

\end{abstract}


\maketitle

\section{Introduction}
With the rapid development of the Internet, accessing multimedia information has become much easier in daily life, and the information overload problem has progressed into a severe challenge. Recommender Systems (RS) have played an important role in helping users pick desired ones from numerous candidates in the fields of social media and e-commerce.
In the field of RS, Collaborative filtering (CF) \cite{MF, NGCF, LightGCN} is one of the traditional algorithms, which uses the historical interaction between users and items to model users' preferences.
CF methods have made great progress in personalized recommendation. In addition, NGCF \cite{NGCF} utilizes Graph Convolutional Networks (GCN) to encode user-item history interactions, and optimize user and item embeddings. LightGCN \cite{LightGCN} modifies the aggregation method of GCN to simplify the model while improving recommendation performance.

However, the development of communication and storage systems makes the item information more diverse, such as text, image, video, etc. The multimodal information of items may have a direct impact on user preferences.
Multimodal Recommender Systems (MRSs), which integrate multimodal information into traditional CF methods, emerge as the times require. Early multimodal recommendations \cite{VBPR, GRCN} regard multimodal data as side information of items, capturing users' preferences. Recent works, such as \cite{DMRL, MARIO}, analyze the user's modality preference through the attention mechanism. Influenced by graph neural network (GNN)-based recommendations \cite{T-MRGF, TMKG, DGCN-HN}, MRSs have also begun to employ GNN to optimize multimodal features. Zhang et al. \cite{LATTICE} perform latent structure learning on visual and textual information to obtain the item' multimodal latent structure, and then optimize the multimodal representation utilizing the GNN. Zhang et al. \cite{MICRO} propagate and aggregate the multimodal information through the GNN, and then obtain the item' multimodal representation by contrastive fusion. Mu et al. \cite{HCGCN} propose a hybrid-clustering-GCN to capture various user behavior patterns to enhance the recommendation performance.

Although the GNN-based MRSs have achieved remarkable results, there are still some limitations. Firstly, most existing works mine the synergistic relationship of items through the item-user-item graph, ignoring the abundant multimodal features that exist in the items. The pretrained embeddings for each modality reflect the modality-specific characteristics, which are absent from the interaction data. Therefore, the item's multimodal semantic relations can be obtained by constructing modality feature graph and item feature graph with pretrained visual and textual features. Secondly, most of the existing MRSs directly use the item multimodal data as side information and analyze the  global impact of multimodal data on user preferences. In fact, the effect of multimodal data on user preferences in different interactions is not invariant. Attention fusion module \cite{WACV} can be utilized to analyze the level of importance of multimodal data in each interaction. Both interaction and multimodal data are regarded as factors affecting user preferences.
In addition, most of the works did not consider the modality differences. It is obvious that the potential feature relationship in multimodal data cannot be fully exploited just by concatenation, because there is a huge semantic gap between visual and textual data naturally. In order to make full use of the multimodal data, we propose a hierarchical fusion network to fuse multimodal features of different granularities. Moreover, we employ contrastive learning to achieve inter-modal alignment of multimodal data, thus eliminating the semantic gap between visual and textual data.

Based on the above analyses, in this paper, we propose a novel aTtention-guided Multi-step FUsion Network for multimodal recommendation, which is termed as \emph{\textbf{TMFUN}} for brevity.
Concretely, in the proposed TMFUN model, we first construct the modality feature graph and the item feature graph to mine the rich multimodal features. 
Then, we design an attention module to explore potential correlations between interaction data and item multimodal features, analyze the effect of multimodal data in different interactions, and achieve early-step fusion of item features. In addition, we design an attention-guided multi-step fusion strategy to fuse item features of different granularities. Finally, we use contrastive learning to achieve the inter-modal alignment of multimodal data.
Our main contributions can be summarized in the following three points:

$\bullet$ We construct the modality feature graph and item feature graph to extract rich multimodal features, and design an attention module to realize the relationship mining between interaction data and multimodal features.

$\bullet$ We design an attention-guided multi-step fusion strategy to achieve multi-granularity fusion of item features, and achieve inter-modal alignment of multi-modal data through contrastive learning.

$\bullet$ The proposed TMFUN outperforms state-of-the-art MRSs, demonstrating the effectiveness of our method.

\section{The Proposed Method}

\begin{figure*}[ht]
	\centering
	\includegraphics[width=0.85\textwidth]{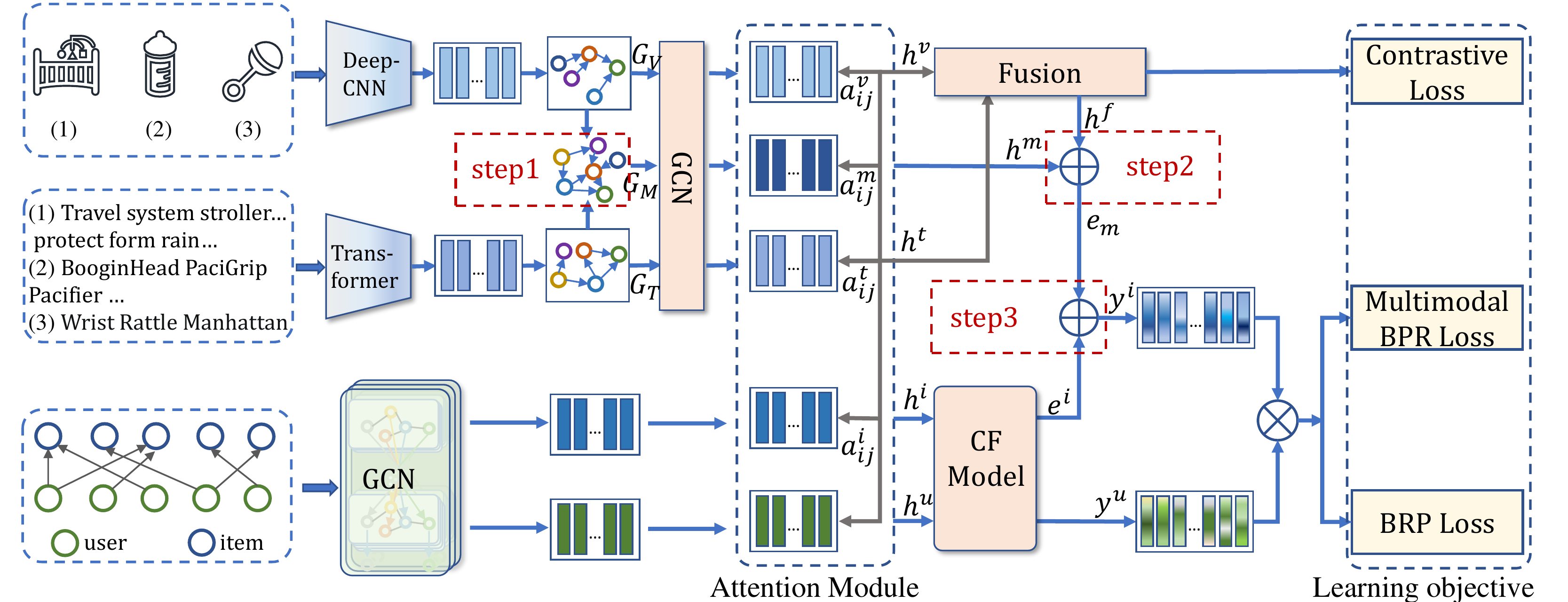}
		\caption{An illustration of the proposed TMFUN model. 
		\label{model}}
\end{figure*}
In this section, we provide a detailed illustration of the proposed model as shown in Figure \ref{model}.

\subsection{Graph Construction and Initialization}
\subsubsection{User-item interaction graph}
In this paper, we define a user-item interaction graph ${{G}_{H}}=\left\{ \left( u,{{y}_{ui}},i \right)\left| \left( u\in U,i\in I \right) \right. \right\}$, where $U$ denotes the user set, $I$ denotes the item set, and ${{y}_{ui}}$ is the connection relationship between user $u$ and item $i$. When there is an explicit relationship between user $u$ and item $i$, ${{y}_{ui}}=1$, otherwise, ${{y}_{ui}}=0$. Following the mainstream recommendation method \cite{LATTICE, KGAT}, we initialize the user and item embeddings in the interaction graph to obtain the ID embeddings ${{x}^{u}},{{x}^{i}}\in {{R}^{d}}$, where $d$ is the dimension.

\subsubsection{Item feature graph}
In this paper, we use the publicly available visual feature extracted by the pretrained convolutional neural network \cite{CNN}, denoted as ${{x}^{v}}\in {{R}^{{{d}_{v}}}}$, where ${d}_{v}=4096$ is the visual feature dimension. To fully mine the associations of image information, we construct the visual feature graph based on the similarity of features. Following \cite{KNN, KNN2},
we use the cosine function to calculate the similarity score between item pair $i$ and $j$,
\begin{equation}
	s_{ij}^{v}=\frac{{{\left( \mathbf{x}_{i}^{v} \right)}^{\top }}\mathbf{x}_{j}^{v}}{\left\| \mathbf{x}_{i}^{v} \right\|\left\| \mathbf{x}_{j}^{v} \right\|},
\end{equation}
where $\mathbf{x}_{i}^{v}$ and $\mathbf{x}_{j}^{v}$ denote the visual feature of item $i$ and $j$ respectively, and $s_{ij}^{v}$ is the similarity score between item pair $i$ and $j$.

Then, for each $i$, we only leave the connections of top-$N$ similar edges to obtain the visual feature graph adjacency matrix ${{S}^{v}}$. The textual feature ${{x}^{t}}\in {{R}^{{{d}_{t}}}}$, ${d}_{t}=378$ is extracted by pretrained sentence-transformers \cite{BERT}. Then, similar to the visual feature graph, we can obtain the textual feature graph adjacency matrix $S^{t}$. 
Finally, we use the freeze fusion strategy to fuse the visual feature graph and text feature graph to obtain the item feature graph adjacency matrix $S$.

\subsection{Attention Relationship Mining}
The message propagation and aggregation mechanism of GCN can optimize the representation of users and items. Therefore, we perform graph convolution operations on user-item interaction graphs, visual feature graph, text feature graph, and item feature graph to obtain the corresponding output${{\widetilde{x}}^{u}}$, ${{\widetilde{x}}^{i}}$, ${{\widetilde{x}}^{v}}$, ${{\widetilde{x}}^{t}}$, ${{\widetilde{x}}^{m}}$.

Then, we design a modality-aware attention mechanism to identify the contribution of each modality $q$ on each user-item interaction. Using ${{\widetilde{x}}^{u}}$ as query, ${{\widetilde{x}}^{i}},{{\widetilde{x}}^{v}},{{\widetilde{x}}^{t}},{{\widetilde{x}}^{m}}$ as key and value, we calculate the effect	$a_{ij}^{q}$ of each modality on each interaction,
\begin{equation}
	a_{ij}^{q}=\frac{\exp \left( \frac{{{\widetilde{x}}^{u}}\odot {{\widetilde{x}}^{q}}}{\sqrt{d}} \right)}{\sum\limits_{q\in Q}{\exp }\left( \frac{{{\widetilde{x}}^{u}}\odot {{\widetilde{x}}^{q}}}{\sqrt{d}} \right)},
\end{equation}	
 where $q\in Q=\{i,v,t,m\}$ represents the modality. $\widetilde{a}_{ij}^{q}=\frac{{{\widetilde{x}}^{u}}\odot {{\widetilde{x}}^{q}}}{\sqrt{d}}$, 
${{\widetilde{x}}^{u}}$ represents the user feature, ${{\widetilde{x}}^{q}}$ is the item feature of the modality $q$, $\odot$ and $d$ denote the dot product and conversion factor respectively. Then, we can obtain the interaction of different modalities, and get the item embedding ${{h}_{ij}}^{q}$ that integrates other modalities,
\begin{equation}
	{{h}_{ij}}^{q}=\sum\limits_{q\in \mathcal{Q}}{a_{ij}^{q}\widetilde{x}_{j}^{q}}.
\end{equation}

\subsection{Multi-step Fusion}
\subsubsection{Modality feature graph fusion}

In order to reduce the model overhead caused by learning fusion coefficients, we adopt the frozen fusion strategy \cite{FREEDOM} to fuse the visual feature graph and textual feature graph to obtain the item-item feature graph ${G}_{M}$.
\begin{equation}
	{{G}_{M}}=\lambda {{G}_{V}}+\left(1-\lambda \right){{G}_{T}},
\end{equation}
where $\lambda$ is the graph fusion coefficient, ${G}_{V}$ is the visual feature graph, and ${G}_{T}$ is the textual feature graph.
\subsubsection{Inter-modal contrastive fusion}
Similarly, we fuse the visual features and textual features, and then use the contrastive loss to supervise the alignment of the visual and textual modalities. 
Then, we can get the item multimodal features ${{e}^{m}}$:
\begin{equation}
	{{e}^{m}}={{h}^{m}}+{{h}^{f}},
\end{equation}
where ${h}^{m}$ is the item multimodal feature obtained from the attention module. ${{h}^{f}}=\mu {{h}^{v}}+(1-\mu ){{h}^{t}}$ is the item fusion feature obtained from contrastive fusion. Where $\mu$ is the contrastive fusion coefficient, ${h}^{v}$ is the visual feature, and ${h}^{t}$ stands for the textual feature.
\subsubsection{Item global fusion}
Then, we fuse the item multimodal features ${e}^{m}$ and the item features ${e}^{i}$ obtained from the CF model \cite{LightGCN} to obtain the final item feature ${{y}^{i}}$,
\begin{equation}
	{{y}^{i}}={{e}^{i}}+{{e}^{m}}.
\end{equation}
\subsubsection{Preference prediction}
In order to predict the user's preference, we first calculate the similarity score between user $u $ and item $i$ according to the inner product,
\begin{equation}
	\hat{y}(u,i)={{\left( {{y}^{u}} \right)}^{\top }}{{y}^{i}},
\end{equation}
where ${y}^{u}$ is the user feature, ${y}^{i}$ it the item feature. Then, the items are sorted according to the similarity scores, and the top $K$ items are recommended to the user, realizing top-$K$ recommendation.

\subsection{Learning Objective}
\subsubsection{Contrastive loss}
With the intention of aligning the modalities, and adaptively capturing the shared feature across multiple modalities, we design a contrastive loss ${\mathcal{L}}_{ {C}}$:

\begin{equation}
{{\mathcal{L}}_{{C}}}=-\frac{1}{\left| I \right|}\sum\limits_{i\in I}{\left[ \frac{1}{\left| Q' \right|}\sum\limits_{q'\in Q'}{\frac{1}{2}}\left( MI\left( h_{i}^{q'},h_{i}^{i} \right)+MI\left( h_{i}^{i},h_{i}^{q'} \right) \right) \right]},
\end{equation}
where $i$ represents item, $I$ denotes the set of items, and $q'\in Q'=\{v,t\}$ denotes the modalities. $MI\left( \cdot,\cdot  \right)$ is the mutual information function, which is used to measure the consistency between features. $h_{i}^{i}$ is the feature of item i, and $h_{i}^{q'}$ denotes the feature of $q'$ modal of item i.

\subsubsection{BPR loss}
We employ BPR loss \cite{BPR} to optimize the model to ensure that the preference score of positive samples is higher than that of negative samples.
\begin{equation}
	{{\mathcal{L}}_{BPR}}=-\sum\limits_{(u,i,j)\in \mathcal{D}}{\log }\sigma \left( {{{\hat{y}}}_{ui}}-{{{\hat{y}}}_{uj}} \right),
\end{equation}
where $D=\{(u,i,j)|(u,i)\in P,(u,j)\in N\}$ is the triplet instance of the training set, $P$ is the positive sample set, $N$ is the negative sample set, and $\sigma \left( \cdot  \right)$ denotes the sigmoid activation function .

\subsubsection{Multimodal BPR loss}
In order to preserve the modality-specific features, we design a multimodal BPR loss to optimize each modality feature of the item,
\begin{equation}
	{{\mathcal{L}}_{mmbpr}}=\sum\limits_{(u,i,j)\in \mathcal{D}}{\sum\limits_{q'\in \{v,t\}}{-}\log \sigma \left( {{\left( {{y}^{u}} \right)}^{\top }}h_{i}^{q'}-{{\left( {{y}^{u}} \right)}^{\top }}h_{j}^{q'} \right)},
\end{equation}
where $D=\{(u,i,j)|(u,i)\in P,(u,j)\in N\}$ is the triplet instance of the training set. $h_{i}^{q'}$ is the multimodal feature of modality $q'\in \{v,t\}$, where $v$ denotes the visual modality and $t$ denotes the text modality. $\sigma \left( \cdot  \right)$ is the sigmoid activation function.
Finally, we can get the learning objective $\mathcal{L}$,
\begin{equation}
	\mathcal{L}={{\mathcal{L}}_{{BPR}}}+\alpha {{\mathcal{L}}_{mmBPR}}+\beta {{\mathcal{L}}_{C}},
\end{equation}
where $\alpha$ is the weight coefficient of multimodal BPR loss ${\mathcal{L}}_{mmBPR}$, and $\beta$ is the weight coefficient of contrastive loss ${\mathcal{L}}_{{C}}$.

\section{EXPERIMENTS}
\subsection{Experimental Setting}

\begin{table}[ht]
  \centering
  \caption{Performance comparison of our TMFUN with other models in terms of Recall@20 (R@20) and NDCG@20 (N@20). The best performance is highlighted in bold.}
   \setlength{\tabcolsep}{0.08cm}{
    \begin{tabular}{l|cc|cc|cc}
    \hline
    \multicolumn{1}{l|}{\multirow{2}{*}{Method}} & \multicolumn{2}{c|}{Baby} & \multicolumn{2}{c|}{Sports} & \multicolumn{2}{c}{Clothing} \\
\cline{2-7}          & R@20 & N@20 & R@20 & N@20 & R@20 & N@20 \\
    \hline
    MF\cite{MF}   & 0.0440 & 0.0200  & 0.0430 & 0.0202 & 0.0191 & 0.0088 \\
    NGCF\cite{NGCF}  & 0.0591 & 0.0261  & 0.0695 & 0.0318 & 0.0387 & 0.0168 \\
    LightGCN\cite{LightGCN} & 0.0698 & 0.0319 & 0.0782 & 0.0369 & 0.0470 & 0.0215 \\
    IMP-GCN\cite{IMP-GCN}  & 0.0720 & 0.0319 & 0.0806 & 0.0378 & 0.0466 & 0.0211 \\
    \hline
    VBPR\cite{VBPR}  & 0.0486 & 0.0213 & 0.0582 & 0.0265 & 0.0481 & 0.0205 \\
    MMGCN\cite{MMGCN} & 0.0640 & 0.0284 & 0.0638 & 0.0279 & 0.0507 & 0.0221 \\
    GRCN\cite{GRCN}  & 0.0754 & 0.0336 & 0.0833 & 0.0377 & 0.0631 & 0.0276 \\
    \hline
    LATTICE\cite{LATTICE} & 0.0829 & 0.0368 & 0.0915 & 0.0424 & 0.0710 & 0.0316 \\
    MICRO\cite{MICRO}   & 0.0865 & 0.0389 & 0.0968 & 0.0445 & 0.0782 & 0.0351\\
    HCGCN\cite{HCGCN}   & 0.0922 & 0.0415 & 0.1032 & 0.0478 & 0.0810 & 0.0370 \\
    \hline
    \textbf{TMFUN}  & \textbf{0.1032} & \textbf{0.0452} & \textbf{0.1112} & \textbf{0.0496} & \textbf{0.0962} & \textbf{0.0431} \\
    \hline
    \end{tabular}}%
  \label{tab1}%
\end{table}%

\subsubsection{Dataset and Parameter Settings}
We conduct experiments on three widely used category subsets of Amazon review dataset \cite{Amazon}: (a) Baby, (b) Sports and Outdoors, and (c) Clothing, Shoes and Jewelry. For simplicity, we represent them using Baby, Sports, and Clothing respectively. 

Following the mainstream methods \cite{LATTICE, HCGCN}, we employ the widely used Recall@K and NDCG@K as our evaluation metrics, referred to as R@K and N@K for short. Specifically, we show the performance when K=20. For each user, we randomly select 80\% of historical interactions for training, 10\% for validation, and the remaining 10\% for testing. Our model is implemented by pytorch and evaluated on an RTX3090 GPU card. We adopt the Xavier method \cite{Xavier} to initialize model parameters and use Adam \cite{Adam} as our optimizer. Furthermore, in our model, the hyperparameters $\lambda$ and $\mu$ are tuned from 0 to 1, and $\alpha$ and $\beta$ are chosen from [1e-4, 1e-3]. In addition, the early stopping and total epochs are set to 20 and 1000 for model convergence.

\subsection{Experimental Result Analysis}
\subsubsection{Performance comparison}

The results of the performance comparison are shown in Table 2. Overall, our proposed TMFUN model outperforms all other models. Concretely, the TMFUN model improves over the state-of-the-art model HCGCN in terms of R@20 by 11.93\%, 7.75\%, and 18.77\% in Baby, Sports, and Clothing, respectively. Compared with existing models, our model further mines the rich multimodal features by constructing modality feature graph and item feature graph, which significantly improves the performance of multimodal recommendation.

\subsubsection{Ablation study}
In this section, as shown in Table \ref{tab2}, the ablation experiments are performed on three datasets. We analyze the influence of the CF method, the attention module, and the multimodal feature, respectively. TMFUN/MF, TMFUN/NGCF, and TMFUN represent different CF methods (MF, NGCF, LightGCN) used in the TMFUN model. TMFUN-mm and TMFUN-att respectively indicate that the multimodal feature and attention module are disabled. We can observe that selecting the appropriate CF method has a significant impact on the model's performance. In the TMFUN model, we adopt LightGCN, which is currently effective and lightweight, as our CF model. In addition, the utilization of multimodal features and the guidance of the attention module have a positive impact on the performance of our model.

\begin{table}[t]
  \centering
  \caption{Performance of our TMFUN model and its variants.}
    \setlength{\tabcolsep}{0.08cm}{
    \begin{tabular}{l|cc|cc|cc}
    \hline
    \multicolumn{1}{l|}{\multirow{2}{*}{Method}} & \multicolumn{2}{c|}{Baby} & \multicolumn{2}{c|}{Sports} & \multicolumn{2}{c}{Clothing} \\
\cline{2-7}          & R@20 & N@20 & R@20 & N@20 & R@20 & N@20 \\
    \hline
    TMFUN/MF      & 0.0852 & 0.0379 & 0.0888 & 0.0381 & 0.0828 & 0.0366 \\
    TMFUN/NGCF    & 0.0812 & 0.0345 & 0.0924 & 0.0397 & 0.0863 & 0.0365 \\
    TMFUN-mm  & 0.0828 & 0.0370 & 0.0913 & 0.0406 & 0.0568 & 0.0252 \\
    TMFUN-att & 0.1022 & 0.0442 & 0.1106 & 0.0489 & 0.0930 & 0.0417 \\
    TMFUN & 0.1032 & 0.0452 & 0.1112 & 0.0496 & 0.0962 & 0.0431 \\
    \hline
    \end{tabular}}%
  \label{tab2}%
\end{table}%

\subsubsection{Parameter analysis}
\begin{figure}[htbp]
	\centering
	\includegraphics[width=0.48\textwidth]{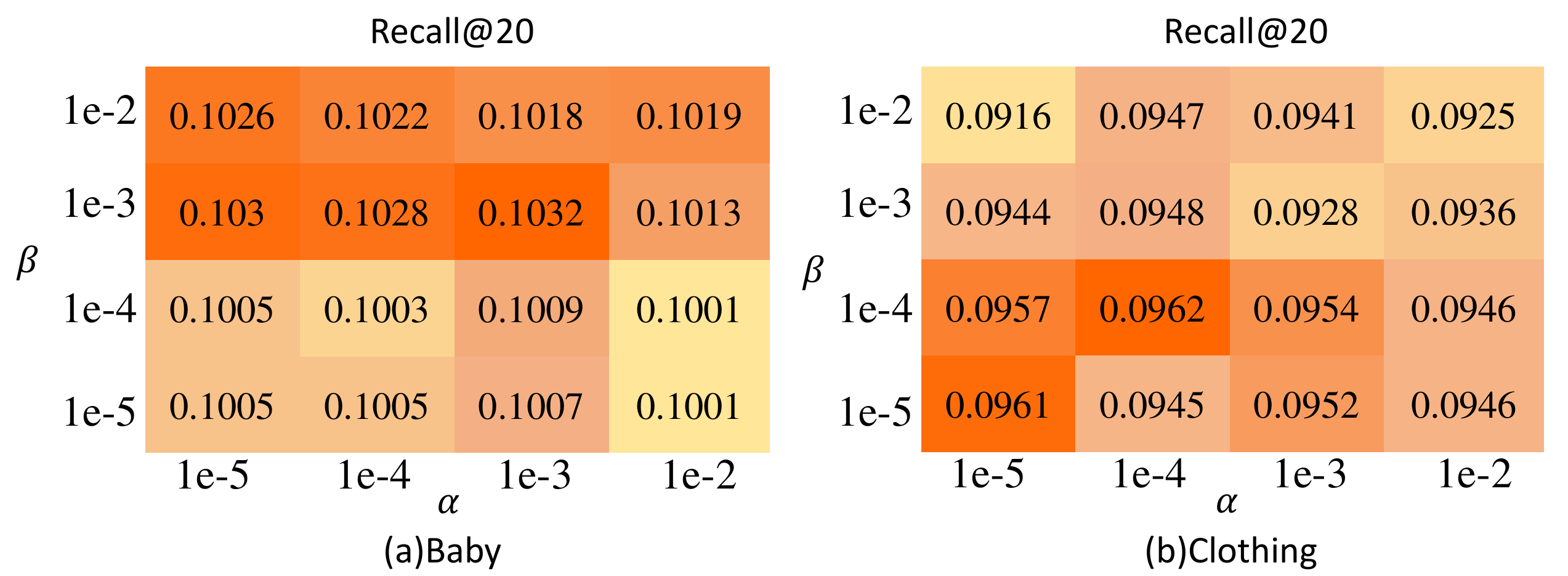}
		\caption{Performance of TMFUN with varying parameters $\alpha$ and $\beta$ in terms of Recall@20 on Baby and Clothing datasets.		
  \label{loss2}}
\end{figure}

In this section, as shown in Figure \ref{loss2}, we carry out experimental verification and analysis on the parameters $\alpha$ and $\beta$ of the learning objective $\mathcal{L}$ to assess their impact.
The parameters of $\alpha$ and $\beta$ are selected from [1e-5, 1e-4, 1e-3, 1e-2]. In Figure 2, we can see that for the Baby dataset, when both $\alpha$ and $\beta$ are 1e-3, the model obtains the optimal performance. For the Clothing dataset, the optimal parameters for $\alpha$ and $\beta$ are 1e-4. Therefore, multimodal BPR loss and contrastive loss play a crucial role in the success of our model, which makes our model fully exploit the latent multimodal information.

\section{CONCLUSION}
To fully exploit the inherent semantic relations contained in the multimodal features, we propose a novel network for multimodal recommendation, named TMFUN.
The core idea of the TMFUN model is to enhance item representation through multimodal data, and provide a more effective preference mining basis for downstream collaborative filtering models. In terms of the feature level, the construction of feature graphs can preserve modality-specific features and the contrastive fusion can capture inter-modal shared features. In terms of the architecture level, we design an attention-guided multi-step fusion strategy to capture the potential relationship between interaction data and multimodal data.
The extensive experiments on three real-world datasets demonstrate the effectiveness of our proposed TMFUN model.
To further mine potential relationships among items, introducing more modalities, such as video and audio, will be implemented in our future work.

\begin{acks}
This work was supported by the National Natural Science Foundation of China (Nos. 62201424 and 62271324), Key Research and Development Program of Shaanxi (Program No. 2023-YBGY-218), and also supported by the ISN State Key Laboratory and High-Performance Computing Platform of Xidian University.
\end{acks}

\bibliographystyle{ACM-Reference-Format}
\bibliography{tmfun}

\end{document}